\documentclass[11pt]{article}
\newcommand{\Comment}[1]{{}}
\usepackage{amsfonts,amsthm,amsmath,amssymb,slashed}
\usepackage[textwidth = 430 pt, textheight = 630 pt]{geometry}
\usepackage{color}

\Comment{\usepackage{color}
\definecolor{MyDarkBlue}{rgb}{0.15,0.15,0.45}
\usepackage[linktocpage=true]{hyperref}
\hypersetup{
colorlinks=true,
citecolor=MyDarkBlue,
linkcolor=MyDarkBlue,
urlcolor=MyDarkBlue,
pdfauthor={Horatiu Nastase},
pdftitle={General $f($) and conformal inflation from minimal supergravity plus matter},
pdfsubject={hep-th}
}

\usepackage[numbers,sort&compress]{natbib}
\usepackage{hypernat}}
\usepackage{graphicx}
\usepackage{cite}

\newcommand\ignore[1]{}
\def\one{{\,\hbox{1\kern-.8mm l}}}

\def\a{\alpha}\def\b{\beta}

\def\d{\partial}

\newcommand{\Cset}{{\,\,{{{^{_{\pmb{\mid}}}}\kern-.45em{\mathrm C}}}}}

\newcommand{\be}{\begin{equation}}
\newcommand{\bea}{\begin{eqnarray}}

\newcommand{\ee}{\end{equation}}
\newcommand{\eea}{\end{eqnarray}}

\begin{document}

\renewcommand{\thefootnote}{\fnsymbol{footnote}}

\makeatletter
\@addtoreset{equation}{section}
\makeatother
\renewcommand{\theequation}{\thesection.\arabic{equation}}

\rightline{}
\rightline{}


\vspace{10pt}


\begin{center}
{\LARGE \bf{\sc General $f(R)$ and conformal inflation from minimal supergravity plus matter}}
\end{center} 
 \vspace{1truecm}
\thispagestyle{empty} \centerline{
{\large \bf {\sc Horatiu Nastase}}\footnote{E-mail address: \Comment{\href{mailto:nastase@ift.unesp.br}}{\tt nastase@ift.unesp.br}}
                                                  }

\vspace{.5cm}


\centerline{{\it 
Instituto de F\'{i}sica Te\'{o}rica, UNESP-Universidade Estadual Paulista}} \centerline{{\it 
R. Dr. Bento T. Ferraz 271, Bl. II, Sao Paulo 01140-070, SP, Brazil}}

\vspace{1truecm}

\thispagestyle{empty}

\centerline{\sc Abstract}

\vspace{.4truecm}

\begin{center}
\begin{minipage}[c]{380pt}
{\noindent We embed general $f(R)$ inflationary models in minimal supergravity plus matter, a single chiral superfield $\Phi$, with or without another superfield
$S$, via a Jordan frame Einstein+scalar description. In particular, inflationary models like a generalized Starobinsky one are analyzed and constraints 
on them are found. We also embed the related models of conformal inflation, also described as Jordan frame Einstein+scalar models,
in particular the conformal inflation from the Higgs model, and analyze the inflationary constraints on them.
}
\end{minipage}
\end{center}

\vspace{.5cm}

\setcounter{page}{0}
\setcounter{tocdepth}{2}

\newpage

\renewcommand{\thefootnote}{\arabic{footnote}}
\setcounter{footnote}{0}

\linespread{1.1}
\parskip 4pt



\section{Introduction}
\ \ \ \ \
Inflation is the leading cosmological model for the initial stages of the evolution of our Universe. On the other hand, one of the best models for 
particle physics at higher energies than the ones we currently probe at accelerators involves supersymmetry. Including gravity in the picture, we 
expect that physics at high energies has as an effective theory given by supergravity. So it is natural to look for inflation in supergravity, and yet obtaining 
good supergravity models of inflation is notoriously difficult, and generally involves some type of fine-tuning. For instance, until recently there were
various negative results ("no-go theorems") for the simplest set-up, for minimal (${\cal N}=1$) supergravity coupled to matter in the form of a single 
chiral superfield, see e.g. \cite{Achucarro:2012hg,Roest:2013aoa}. Recently however, models embedding rather general potentials within ${\cal N}=1$ 
supergravity with one chiral superfield were proposed \cite{Ketov:2014qha,Ketov:2014hya}, as well as ways to embed general potentials within 
${\cal N}=1$ supergravity with one chiral superfield ($\Phi$), plus another one ($S$) stabilized at zero \cite{Roest:2013aoa}. 
Also, a special class of models that 
has been called $\a$-attractors can be embedded in ${\cal N}=1$ supergravity with one chiral superfield \cite{Roest:2015qya,Linde:2015uga}.

In this paper we are interested in the embedding in minimal supergravity plus matter of Jordan frame Einstein+scalar models, which can be written 
as Einstein+scalar with a potential. One such class of models are the $f(R)$ models. We will show that a generic $f(R)$ model can be written as a
Einstein+scalar model, and reversely, any Einstein+scalar with potential model can be written in $f(R)$ form. In particular, we will analyze several 
inflationary potentials from the point of view of $f(R)$ and of the minimal supergravity embedding. In the constraints, we will use the results of the 
Planck \cite{Planck:2013jfk} and WMAP \cite{wmap9} experiments, but not the value of the tensor to scalar ratio $r$ from BICEP2 \cite{Ade:2014xna},
since there is uncertainty surrounding it \cite{Flauger:2014qra,Mortonson:2014bja} and recently the joint Planck and BICEP2 paper 
drastically modified the result \cite{Ade:2015tva}.

We will also consider another class of Jordan frame Einstein+scalar 
models that goes under the name of conformal inflation. These are models with two scalar fields and 
local Weyl symmetry, found in \cite{Kallosh:2013hoa,Kallosh:2013maa}, following earlier work by 
\cite{Ferrara:2010in,Bars:2011mh,Bars:2011aa,Bars:2012mt,Kallosh:2013lkr,Kallosh:2013pby}. These models generically give rise to the 
same predictions as the Starobinsky model \cite{Starobinsky:1980te}, since the asymptotic Einstein-frame 
scalar potential in the inflationary region is the same. In \cite{Costa:2014lta} it was considered the possibility that the inflaton is also the Higgs,
since by now the Higgs is the only discovered scalar \cite{Aad:2012tfa,Chatrchyan:2012ufa},
and moreover it was found that with some rather unusual choices 
for an arbitrary function we can get a generalized type of Starobinsky model in the inflationary region 
(the idea of Higgs inflation has a long history; for the present discussion we note that the Bezrukov-Shaposhnikov model \cite{Bezrukov:2007ep}
admits a Weyl-symmetric formulation \cite{Bars:2013yba}). 
In this paper we will see that we can actually get any potential of the "new inflation" type, and we will investigate the embedding in supergravity 
of these conformal Higgs inflation models, and their relation to $f(R)$ models. Note that the generalized Starobinsky model was considered before, 
for instance in \cite{Ellis:2013nxa,Ellis:2014cma}, and in the context of supergravity with two chiral superfields in 
\cite{Ferrara:2013rsa,Kallosh:2013yoa,Cecotti:2014ipa}.
After the paper first appeared on arXiv, I became aware of other papers dealing with issues related to the ones described in this paper:
in \cite{Ozkan:2014cua} it was considered a supersymmetrization of $R+R^n$ Starobinsky-like models, in \cite{Ozkan:2015iva} it was shown that 
the $\a$-attractors later embedded in supergravity in  \cite{Roest:2015qya,Linde:2015uga} can also be related to $f(R)$ models coupled to an 
auxiliary vector field, and in \cite{Broy:2014xwa} it was analyzed the relation between $f(R)$ models and generalized versions of the Starobinsky 
model.

The paper is organized as follows. In section 2 we will first show that a general $f(R)$ action can be obtained from Einstein-Hilbert plus a dynamical 
scalar, and then embed them in minimal supergravity. In section 3 we will focus on examples relevant for inflation and consider inflationary constraints
on them. In section 4 we change the focus to conformal inflation models, and show how to embed them in minimal supergravity. In section 5 we 
consider the set-up of conformal inflation coming from the Higgs, inflationary models related to it, and inflationary constraints on them, and in 
section 6 we conclude.

\section{General $f(R)$ from minimal supergravity}

\subsection{f(R) actions as EH plus dynamical scalar}

There is a general procedure for writing an $f(R)$ action as a usual Einstein-Hilbert one plus a dynamical scalar field. For instance, in the case of a 
monomial correction to the Einstein-Hilbert action, it was described e.g. in \cite{Costa:2014lta}. One writes a first order form for the $f(R)$ 
action, in terms of an action linear in $R$, with an auxiliary scalar. When going to the Einstein frame, the auxiliary scalar becomes dynamical, and 
acquires a nontrivial potential. The procedure is not unique (though of course the result written in terms of a canonical scalar $\varphi$ is).

We start with a slightly different construction for the $R+R^{p+1}$ action than in \cite{Costa:2014lta}, 
which is easier to generalize. It is easier to start with the action linear in $R$ 
and with an auxiliary scalar $\a$,
\be
S=\frac{M_{\rm Pl}^2}{2}\int d^4x \sqrt{-g}\; [R(1+\a^p)-\b\a^q].
\ee
Varying with respect to $\a$, we obtain
\be
R=\b\frac{q}{p}\a^{q-p}\Rightarrow \a=\left[\frac{p}{q\b } R\right]^{\frac{1}{q-p}}\;,
\ee
and substituting back in the action we obtain 
\be
S=\frac{M_{\rm Pl}^2}{2}\int d^4x \sqrt{-g}\; \left\{R+\frac{R^{\frac{q}{q-p}}}{\b^{\frac{p}{q-p}}}\left[\left(\frac{p}{q}\right)^{\frac{p}{q-p}}
-\left(\frac{p}{q}\right)^{\frac{q}{q-p}}\right]\right\}.
\ee
It is clear that by varying $\b$ and $q$ and $p$, we can obtain any coefficient and power for the monomial correction.

In particular, a case that would be experimentally favored for the present day Universe, with the monomial being approximately a cosmological constant, i.e.
$q=\epsilon\ll 1$ (and $p=1$) gives
\be
S=\int d^4x \sqrt{-g}\; [R(1+\a)-\b \a^\epsilon]\leftrightarrow S\simeq \int d^4x\sqrt{-g}\; [R-\b R^{-\epsilon}].\label{frepsilon}
\ee
However, as we shall shortly see, this does not give a good inflationary potential.

The equivalence to an Einstein-Hilbert action plus dynamical scalar is obtained by first defining the Einstein metric
\be
g_{\mu\nu}^E=[1+\a^p]g_{\mu\nu}\equiv \Omega^{-2}g_{\mu\nu}\;,\label{einsteinjordan}
\ee
and then using the general formula for a Weyl rescaling in $d$ dimensions
\be
R [g_{\mu\nu}]=\Omega^{-2}\left[R[g^E_{\mu\nu}]-2(d-1)g^{\mu\nu}_E\nabla^E_\mu\nabla^E_\nu\ln \Omega-(d-2)(d-1)g^{\mu\nu}_E(\nabla_\mu^E\ln\Omega)
\nabla_\nu^E\ln\Omega\right]\;,
\ee
thus obtaining the Einstein plus scalar action
\be
S=\frac{M_{\rm Pl}^2}{2}\int d^4x\sqrt{-g_E}\left[R[g_E]-\frac{3}{2}g^{\mu\nu}_E\frac{\nabla_\mu^E\a^p\nabla_\nu^E\a^p}{[1+\a^p]^2}\label{E+scalar}
-2V(\a)\right]\;,
\ee
where the potential is given by
\be
2V(\a)=\frac{\beta\a^q}{\left[1+\a^p\right]^2}\;,
\ee
and the canonical scalar $\varphi$ is defined by 
\be
\a^p=e^{\sqrt{\frac{2}{3}}\frac{\varphi}{M_{\rm Pl}}}-1.
\ee

Note that specializing to the case $p=1, q=\epsilon\ll 1$, we obtain 
\be
2V(\varphi)=\b \left(e^{\sqrt{\frac{2}{3}}\frac{\varphi}{M_{\rm Pl}}}-1\right)^\epsilon e^{-2\sqrt{\frac{2}{3}}\frac{\varphi}{M_{\rm Pl}}}\;,\label{epsilonpot}
\ee
which it's clear that doesn't give  a good inflationary potential, despite the naive expectation based on the form of the $f(R)$ action (\ref{frepsilon}). 

We can use the case $p=1$ and generalize the construction to an action with a general 
$f(R)$ correction to the Einstein-Hilbert term. Again starting with the action with an auxiliary field
\be
S=\frac{M_{\rm Pl}^2}{2}\int d^4x\sqrt{-g}\; [R(1+\a)-\b g(\a)]\;,
\ee 
we can solve for $\a$, giving $R=\b g'(\a)$, or
\be
\a=(g')^{-1}\left(\frac{R}{\b}\right)\;,
\ee
and substituting in the action we get
\be
S=\frac{M_{\rm Pl}^2}{2}\int d^4x\sqrt{-g}\; \left[R\left(1+(g')^{-1}\left(\frac{R}{\b}\right)\right)-\b g\left((g')^{-1}\left(\frac{R}{\b}\right)\right)\right]\;,
\ee
which means that the action describes an $f(R)$ model with 
\be
f(R)=R(g')^{-1}\left(\frac{R}{\b}\right)-\b g\left((g')^{-1}\left(\frac{R}{\b}\right)\right).\label{fg}
\ee
Via the same change to the Einstein metric (\ref{einsteinjordan}), the action is again (\ref{E+scalar}), but now with $p=1$ and 
\be
2V=\frac{\b g(\a)}{(1+\a)^2}=\b e^{-2\sqrt{\frac{2}{3}}\frac{\varphi}{M_{\rm Pl}}}g\left(e^{\sqrt{\frac{2}{3}}\frac{\varphi}{M_{\rm Pl}}}-1\right)\;,\label{potfr}
\ee
and a canonical scalar defined by $\a=e^{\sqrt{\frac{2}{3}}\frac{\varphi}{M_{\rm Pl}}}-1$.

It seems that any function $f(R)$ should be describable in terms of some function $g(R)$ via (\ref{fg}), though I have not been able to prove it.
What is indeed clear is that reversely, any potential $V(\varphi)$ can be derived from a $g(X)$, given by
\be
g(X)=\frac{(X+1)^2}{\b}V\left(M_{\rm Pl}\sqrt{\frac{3}{2}}\ln (X+1)\right)\;
\ee
and thus from an $f(R)$ given by (\ref{fg}).

In particular, a generalized Starobinsky model was defined in \cite{Costa:2014lta} implicitly, via the action 
\be
S=\frac{M_{\rm Pl}^2}{2}\int d^4x\sqrt{-g}\left[R(1+\a)^b-\b\a^{2b}\right]\;,
\ee
leading to the potential 
\be
2V=\b\frac{(\a)^{2b}}{[1+\a]^{2b}}=\b\left[1-e^{-\sqrt{\frac{2}{3}}\frac{\phi}{bM_{\rm Pl}}}\right]^{2b}
\simeq \b\left[1-2be^{-\sqrt{\frac{2}{3}}\frac{\phi}{bM_{\rm Pl}}}\right]\;,\label{genstarob}
\ee
where the approximation is for $\varphi\rightarrow\infty$, 
and the canonical scalar is defined by $\a=e^{\sqrt{\frac{2}{3}}\frac{\phi}{bM_{\rm Pl}}}-1$.
We could now write down explicitly $g(X)$, though not $f(R)$.

\subsection{Embeddings in minimal supergravity}

{\bf Single chiral superfield}

In \cite{Ketov:2014hya} (see also \cite{Ketov:2014qha}), a way to embed general inflationary potentials in minimal supergravity plus a single chiral 
superfield was described. We will see that it can be applied to our case. Consider the K\"{a}hler potential (we put $M_{\rm Pl}=1$ for simplicity
whenever we consider superfields)
\be
K=-3\ln\left[1+\frac{\Phi+\bar\Phi}{\sqrt{3}}\right]\;,
\ee
and consider the canonical inflaton $\varphi$ to be the imaginary part of $\Phi$, while the real part is stabilized at zero, $\langle {\rm Re}\Phi\rangle=0$,
i.e.
\be
\varphi=\sqrt{2}{\rm Im} \Phi.
\ee
Then the kinetic term of $\Phi$ is canonical, 
\be
{\cal L}=-\d_\mu\bar \Phi\d^\mu\Phi\;,
\ee
and then the general ${\cal N}=1$ supergravity plus chiral superfield potential formula,
\be
V=e^K\left[g^{\Phi\bar\Phi}|D_\Phi W|^2-3|W|^2\right]\;,
\ee
with $D_\Phi W=\d_\Phi W+(\d_\Phi K)W$, becomes simply 
\be
V(\varphi)=|\d_\Phi W(i{\rm Im}\Phi|^2=(\hat W'(\varphi))^2\;,
\ee
where 
\be
W(\Phi)=\frac{1}{\sqrt{2}}\hat W(-\sqrt{2}i\Phi)\;,
\ee
{\em if $\hat W$ is a real function of its argument}.

For the generalized Starobinsky model (\ref{genstarob}), we get the derivative of $\hat W$
\be
\hat W'(\varphi)=\sqrt{\b/2}\left[1-e^{-\sqrt{\frac{2}{3}}\frac{\varphi}{b}}\right]^b\;,
\ee
which gives the superpotential
\be
W(\Phi)=\frac{\sqrt{3\b/2}}{2}(-1)^{1-b}e^{-\sqrt{\frac{2}{3}}\Phi}{}_2F_1(-b,-b,1-b;e^{\sqrt{\frac{2}{3}}\frac{\Phi}{b}}).\label{wgenstarob}
\ee

For the general $f(R)$ in (\ref{fg}), with scalar potential (\ref{potfr}), we obtain
\be
\hat W'(\varphi)=\sqrt{\b/2}e^{-\sqrt{\frac{2}{3}}\varphi}\sqrt{g\left(e^{\sqrt{\frac{2}{3}}\varphi}-1\right)}\;,
\ee
leading to the superpotential
\be
W(\Phi)=\sqrt{\b}\int d\Phi e^{-\sqrt{\frac{2}{3}}\Phi}\sqrt{g\left(e^{\sqrt{\frac{2}{3}}\Phi}-1\right)}.\label{wintg}
\ee

{\bf Two chiral fields}

In \cite{Roest:2013aoa}, another way to embed inflationary models in minimal supergravity was considered, with two superfields
$\Phi$ and $S$, but the second being stabilized at zero, $\langle S\rangle =0$. The K\"{a}hler potential is 
\be
K=-\a\log(\Phi+\bar \Phi -S\bar S)\;,
\ee
and the superpotential is 
\be
W=Sf(\Phi).
\ee
From the general ${\cal N}=1$ supergravity formula
\be
V=e^K\left[g^{i\bar j}(D_iW)D_{\bar j}\bar W-3|W|^2\right]\;,
\ee
with $X\equiv \Phi+\bar\Phi-S\bar S$, we get at $S=0$, 
\be
V(S=0)=e^Kg^{S\bar S}|D_SW|^2=\frac{X^{1-\a}}{\a}|f|^2.\label{fphis}
\ee
We see that we have to take $R={\rm Re}\Phi$ as the inflaton, with the canonical inflaton $\varphi$ being found from 
$g_{\Phi\bar\Phi}=\a/(4R^2)$ as 
\be
R=e^{\sqrt{\frac{2}{\a}}\varphi}.
\ee

For the generalized Starobinsky model, equating (\ref{genstarob}) with (\ref{fphis}), we obtain the superpotential
\be
W(\Phi,S)=S2^{\frac{\a-3}{2}}\sqrt{\a\b}\Phi^{\frac{\a-1}{2}}\left(1-\Phi^{-\sqrt{\frac{\a}{3}}\frac{1}{b}}\right)^b.\label{wsgenstarob}
\ee
The Starobinsky case $\a=3,b=1$, for the coefficient $\b=3$, gives
\be
W=3S(\Phi-1).
\ee
Note however that from (\ref{fphis}), various $\a$'s give various potentials for Im$\Phi$ corresponding to the same $V(\varphi)$. 

For the general $f(R)$ defined by the function $g(R)$, the potential (\ref{potfr}) equated with (\ref{fphis}) gives the superpotential
\be
W=S\sqrt{\a\b/2}(2\Phi)^{\frac{\a-1}{2}}\Phi^{-\sqrt{\frac{\a}{3}}}\sqrt{g\left(\Phi^{\sqrt{\frac{\a}{3}}}-1\right)}.
\ee
For $\a=3$, the form becomes simpler,
\be
W=\sqrt{6\b}S\sqrt{g(\Phi-1)}.\label{wg}
\ee
Considering the monomial function, 
\be
g(X)=\frac{X^a}{a}\;,
\ee
giving 
\be
f(R)=\frac{a-1}{a}\frac{R^{\frac{a}{a-1}}}{\b^{\frac{1}{a-1}}}\;,
\ee
the superpotential is 
\be
W(\Phi,S)=\sqrt{6\b}S\frac{(\Phi-1)^{a/2}}{\sqrt{a}}.
\ee

\section{Inflationary $f(R)$ models and constraints}

We can now specialize to inflationary models of the $f(R)$ type. We have already seen the superpotentials (\ref{wgenstarob}) and 
(\ref{wsgenstarob}) for the generalized Starobinsky model. Another possibility is to take a potential that is simply
\be
V=\b\left[1-ce^{-\sqrt{\frac{2}{3}}\frac{\a\varphi}{M_{\rm Pl}}}\right].
\ee
This corresponds to the function 
\be
g(X)=(X+1)^2[1-c(X+1)^{-\a}]\;,
\ee
which via (\ref{wg}) leads to the two chiral superfield superpotential
\be
W(\Phi,S)=\sqrt{6\b}S\Phi\sqrt{1-c\Phi^{-\a}}\;,
\ee
and via (\ref{wintg}) leads to the single chiral superfield superpotential
\be
W(\Phi)=\sqrt{\frac{\b}{2}}\left\{-\frac{\sqrt{6}}{\a}\sqrt{1-ce^{-\sqrt{\frac{2}{3}}\a\Phi}}+\Phi +\frac{\sqrt{6}}{\a}\log\left[
1+\sqrt{1-ce^{-\sqrt{\frac{2}{3}}\a\Phi}}\right]\right\}.
\ee

The $W(\Phi,S)$ superpotential can be thought of as summing an infinite series of quantum corrections to the classical $W\sim S\Phi$ piece.

As explained in \cite{Costa:2014lta}, defining the usual inflationary parameters
\bea
\epsilon&=&\frac{M_{\rm Pl}^2}{2}\left(\frac{V'(\varphi)}{V(\varphi)}\right)^2\cr
\eta&=&\frac{M_{\rm Pl}^2 V''(\varphi)}{V(\varphi)}\cr
N_e&=&-\int_{\varphi_0}^{\varphi_f}\frac{d\varphi/M_{\rm Pl}}{\sqrt{2\epsilon}}\;,
\eea
we find that $\epsilon\ll \eta$ and 
\be
n_s-1\simeq 2\eta;\;\;\;\;
r=16\epsilon=\frac{2}{\a^2}(n_s-1)^2;\;\;\;\;
N_e=\frac{2}{1-n_s}.
\ee
For example, $N_e=50$ gives $n_s=0.9600$ and $N_e=60$ gives $n_s=0.9667$, both compatible with the Planck+WMAP result 
$0.9603\pm 0.0073$ \cite{Planck:2013jfk}. Since there is uncertainty in the determination of $r$, the initial result of BICEP2 \cite{Ade:2014xna} being 
modified, we will not consider $r$ for excluding models.

The generalized Starobinsky model (\ref{genstarob}) corresponds to the function
\be
g(X)=(X+1)^2[1-(X+1)^{-\frac{1}{b}}]^{2b}\;,
\ee
so in particular the Starobinsky model, with $b=1$, corresponds to $g(X)=X^2$. The analysis at large $\varphi$ is the same as the above, for
$\a=1/b$. The constraint on $n_s$ is the same, and only on $r$ is modified, but as we said we will not consider it for the purpose of excluding models.

As we saw, we can derive any potential from an $g(X)$, and thus from an $f(R)$. In particular, for the usual power law chaotic inflation, 
with 
\be
V=\lambda_p\phi^p\;,\label{phip}
\ee
which gives as usual
\bea
\epsilon&=&\frac{p^2}{2}\left(\frac{M_{\rm Pl}}{\varphi_*}\right)^2\cr
\eta&=&p(p-1)\left(\frac{M_{\rm Pl}}{\varphi_*}\right)^2\cr
N_{e,*}&\simeq &\frac{1}{2p}\left(\frac{\varphi_*}{M_{\rm Pl}}\right)^2\;,
\eea
so 
\be
n_s-1=-6\epsilon+2\eta=-\frac{p+2}{2N_{e,*}};\;\;\;
r=16\epsilon=\frac{4p}{N_{e,*}}\;,
\ee
we obtain the function 
\be
g(X)=(X+1)^2\ln^p(X+1)\;,
\ee
which for general $p$ has a derivative $g'(X)=p(X+1)\ln^{p-1}(X+1)+2(X+1)\ln^p(X+1)$ 
that is not so easy to invert in order to obtain $f(R)$ explicitly. 

For the simplest model, $p=1$, $g'(X)$ can be inverted only approximately. For $X+1$ very small or very large, we have 
\be
X+1+2(X+1)\ln(X+1)=\pm y\Rightarrow X\simeq -1+\frac{\pm y}{1+2\ln (y)}\;,
\ee
so 
\be
f(R)\simeq -R+\frac{R^2}{\b(1+2\ln(R/\b))^2}\left[\ln(R/\b)+\ln(2\ln(R/\b))\right]\;,
\ee
and the Einstein-Hilbert term cancels. We see that in the $f(R)$ picture, this simplest of models, EH term plus a free scalar with a linear potential, 
already looks quite complicated. 

The class of monomial chaotic inflation potentials (\ref{phip}) can be embedded in supergravity via (\ref{wg}) with the superpotential
\be
W(\Phi,S)=\sqrt{6\b}S\Phi\ln^{p/2}\Phi\;,
\ee
and via (\ref{wintg}) with the superpotential
\be
W=\frac{(2/3)^{p/4}\sqrt{\lambda_p}}{p/2+1}\Phi^{p/2+1}\;,
\ee
so even powers $p$ are obtained from simple polynomial superpotentials.

The chaotic inflation monomials are again viable experimentally, $p=2$ giving $1-n_s=2/N_e$ as the Starobinsky model, only with a different $r$, $r=8/N_e$.

Other examples of inflationary potentials can be treated similarly.

\section{Conformal inflation from minimal supergravity}

Models of conformal inflation are models with two real scalars and a local Weyl symmetry, as well as a $SO(1,1)$ invariance at large field values. 
They have been defined in \cite{Kallosh:2013hoa,Kallosh:2013maa}. The extension described here was found in \cite{Costa:2014lta}.

One starts with the Einstein-scalar action in a Jordan frame
\be
S=\frac{1}{2}\int d^4x \sqrt{-g}\; \left[\d_\mu\chi\d^\mu\chi-\d_\mu\phi\d^\mu\phi+\frac{\chi^2-\phi^2}{6}R-2V\right]\;,
\ee
where both the kinetic terms and the potential $V$ are invariant under the local Weyl symmetry
\be
g_{\mu\nu}\rightarrow e^{-2\sigma(x)}g_{\mu\nu};\;\;\;\;
\chi\rightarrow e^{\sigma(x)}\chi;\;\;\;\;
\phi\rightarrow e^{\sigma(x)}\phi\;,
\ee
allowing the elimination of one of the scalars through gauge fixing. In particular, a gauge choice (Einstein gauge) leading to the Einstein frame is 
\be
\chi=\sqrt{6}M_{\rm Pl}\cosh\frac{\varphi}{\sqrt{6}M_{\rm Pl}};\;\;\;\;
\phi=\sqrt{6}M_{\rm Pl}\sinh\frac{\varphi}{\sqrt{6}M_{\rm Pl}}.
\ee

The kinetic terms are invariant under the $SO(1,1)$  symmetry acting on 
$(\chi,\phi)$, and one imposes the same symmetry at large $\chi$ and $\phi$ for the potential $V$. This fixes 
\be
V=\lambda f(\phi/\chi)[\phi^2-h(\phi/\chi)\chi^2]^2\;,
\ee
with $h(1)=1$.

Considering the models with $f\equiv 1$ in Einstein gauge, the Einstein frame potential is
\be
V=36\lambda M_{\rm Pl}^4\left[\sinh^2\frac{\varphi}{\sqrt{6}M_{\rm Pl}}-h\left(\tanh\frac{\varphi}{\sqrt{6}M_{\rm Pl}}\right)\cosh^2\frac{\varphi}{\sqrt{6}
M_{\rm Pl}}\right]^2.
\ee

Then, using the embedding with a single chiral superfield, we obtain the superpotential
\be
W(\Phi)=3\sqrt{\lambda}M_{\rm Pl}^2\int d\Phi \left[\sinh^2\frac{\Phi}{\sqrt{6}M_{\rm Pl}}
-h\left(\tanh\frac{\Phi}{\sqrt{6}M_{\rm Pl}}\right)\cosh^2\frac{\Phi}{\sqrt{6}M_{\rm Pl}}\right]\;,
\ee
and using the embedding with two chiral superfields, we obtain the superpotential
\bea
W(\Phi,S)&=&6\sqrt{\lambda\alpha}M_{\rm Pl}^2S(2\Phi)^{\frac{\a-1}{2}}\left[\frac{(\Phi/M_{\rm Pl})^{\sqrt{\frac{\a}{3}}}+(\Phi/M_{\rm Pl})
^{-\sqrt{\frac{\a}{3}}}-2}{4}\right.\cr
&&\left.-h\left(\frac{(\Phi/M_{\rm Pl})^{\frac{1}{2}\sqrt{\frac{\a}{3}}}-(\Phi/M_{\rm Pl})^{-\frac{1}{2}\sqrt{\frac{\a}{3}}}}{(\Phi/M_{\rm Pl})^{\frac{1}{2}\sqrt{\frac{\a}{3}}}
+(\Phi/M_{\rm Pl})^{-\frac{1}{2}\sqrt{\frac{\a}{3}}}}\right)\frac{(\Phi/M_{\rm Pl})^{\sqrt{\frac{\a}{3}}}+(\Phi/M_{\rm Pl})^{-\sqrt{\frac{\a}{3}}}+2}{4}\right].\cr
&&
\eea
For $\a=3$, we obtain
\be
W(\Phi,S)=12\sqrt{3\lambda}M_{\rm Pl}^3S\left[\frac{(\Phi/M_{\rm Pl}-1)^2}{4}-h\left[\frac{\Phi/M_{\rm Pl}-1}{\Phi/M_{\rm Pl}+1}\right]
\frac{(\Phi/M_{\rm Pl}+1)^2}{4}\right].
\ee

We see that in $W(\Phi)$, at small $\Phi$ we have a renormalizable superpotential, 
\be
W(\Phi)\simeq \sqrt{\lambda/2}\int d\Phi\left[\Phi^2-h\left(\frac{\Phi}{\sqrt{6}M_{\rm Pl}}\right)\left(6M_{\rm Pl}^2+\Phi^2\right)\right]\;,
\ee
so we can think of the large $\Phi$ function as a UV completion. In $W(\Phi,S)$, for $\a=3$, we have again a 
renormalizable superpotential.

\section{Conformal inflation from the Higgs models and constraints}

As explained in \cite{Costa:2014lta}, the conformal inflaton $\varphi$ can be thought to be related to the Higgs, 
specifically by $\phi=\sqrt{H^\dagger H}$. In that case we want to obtain the Higgs potential at small energies, which means
we need $h(0)\simeq \omega^2$, where $\omega\simeq 246 GeV/\sqrt{6}M_{\rm Pl}$, and the function $h(x)$ must give subleading 
corrections at small energies, so in the simplest case, of a polynomial plus a constant, thus
\be
h(x)=\omega^2+(1-\omega^2)x^n\;,\label{hofx}
\ee
with $n>2$. Then, at small $x$, we get
\be
V\simeq \lambda[\varphi^2-(246GeV)^2]^2\;,
\ee
plus higher corrections, as we want. With the above $h(x)$, the superpotential $W(\Phi)$ at small field, including the first correction, is 
\be
W(\Phi)\simeq\sqrt{\lambda}/2(\sqrt{6}M_{\rm Pl})^3\left[\frac{1}{3}\left(\frac{\Phi}{\sqrt{6}M_{\rm Pl}}\right)^3-\omega^2\frac{\Phi}{\sqrt{6}M_{\rm Pl}}
-\frac{1}{n+1}\left(\frac{\Phi}{\sqrt{6}M_{\rm Pl}}\right)^{n+1}\right]\;,
\ee
and the superpotential $W(\Phi,S)$, including the first correction, is 
\be
W(\Phi,S)\simeq 12\sqrt{\a\lambda}M_{\rm Pl}^3\Phi S\left[\frac{(\Phi/M_{\rm Pl}-1)^2}{4}-\omega^2\frac{(\Phi/M_{\rm Pl}+1)^2}{4}
-\frac{1}{4}\frac{(\Phi/M_{\rm Pl}-1)^{n+2}}{(\Phi/M_{\rm Pl}+1)^n}\right]\;,
\ee
and one could easily see such superpotentials for the Higgs field appearing in specific models. 

For generic embedding functions $f(x)$ and $h(x)$ (that are not too singular), one obtains the Starobinsky model, for instance for (\ref{hofx}) one obtains at 
$\varphi\rightarrow \infty$, 
\be
V(\varphi)\simeq 9(n-2)^2\lambda M_{\rm Pl}^4\left[1-2n e^{-\sqrt{\frac{2}{3}}\frac{\varphi}{M_{\rm Pl}}}\right].\label{genericstarob}
\ee
If the functions admit a Taylor expansion at $x=1$, we will obtain the Starobinsky model, so we must look for functions that have a different behaviour 
there.

In order to avoid the Starobinsky model by having a faster deviation than $e^{-\sqrt{\frac{2}{3}}\frac{\varphi}{M_{\rm Pl}}}$
from the asymptotic constant potential, it is easier to construct functions $f(x)$, for the same $h(x)$ as above.
For example, one way to generate a generalized Starobinsky model is via a function 
\be
f(x)=1+C\left[\ln\frac{2}{1+x^p}\right]^\a\;,
\ee
with $p>2$. Indeed, then at $\varphi\rightarrow \infty$ we obtain 
\be
f(x)\simeq 1+C p^\a e^{-\sqrt{\frac{2}{3}}\frac{\a\varphi}{M_{\rm Pl}}}\;,
\ee
and for $\varphi\rightarrow 0$ we get
\be
f(x)\simeq 1+C\ln^\a 2 -\a C\ln^{\a-1}2 \left(\frac{\varphi}{\sqrt{6} M_{\rm Pl}}\right)^p.
\ee
Then, if $\a<1$, the deviation from the asymptotic constant at $\varphi\rightarrow \infty$ is larger than the deviation due to $h(x)$, hence it dominates. 
We see now that the more precise condition on the function $f(x)$ is $f(1-x)-1\propto x^{\a_1}$ and $f(x)-1\sim c_1+c_2x^p$, with $\a_1\neq 1$ and
$p>2$, for $x\rightarrow 0$. 

We can also generate an inflationary potential of the type
\be
V=V_0\left(1-K\left(\frac{M_{\rm Pl}}{\varphi}\right)^p\right)\;,\label{v0power}
\ee
which lead to the slow-roll inflationary parameters
\bea
\epsilon&=&\frac{1}{2}\left(pK\left(\frac{M_{\rm Pl}}{\varphi_*}\right)^{p+1}\right)^2\ll |\eta|\cr
\eta&=&-(p+1)K\left(\frac{M_{\rm Pl}}{\varphi_*}\right)^{p+2}\cr
N_{e,*}&=&\frac{1}{p(p+2)K}\left(\frac{\varphi_*}{M_{\rm Pl}}\right)^{p+2}\;,
\eea
and thus to the observables
\bea
1-n_s&\simeq&-2\eta=\frac{2(p+1)}{p(p+2)}\frac{1}{N_{e,*}}\cr
r&=&16\epsilon=\frac{8p^2 K^{\frac{2}{p+2}}}{[p(p+2)]^{\frac{2p+2}{p+2}}}\frac{1}{N_{e,*}^{\frac{2p+2}{p+2}}}.
\eea
Such inflationary potentials are obtained for instance from the function 
\be
f(x)=1-C\ln^{-p}\left(\frac{1-x^q}{2q}\right)\;,
\ee
with $q>2$, which gives at $\varphi\rightarrow\infty$
\be
f(x)\simeq 1-C\left(-\sqrt{\frac{3}{2}}\right)^p\left(\frac{M_{\rm Pl}}{\varphi}\right)^p\;,
\ee
and at $\varphi\rightarrow 0$
\be
f(x)\simeq 1-C(-\ln(2q))^{-p}\left(1-\frac{p}{\ln (2q)}\left(\frac{\varphi}{\sqrt{6}M_{\rm Pl}}\right)^q\right).
\ee
Since the power law is a larger deviation from the asymptotic constant than the exponential, the correction at $\varphi\rightarrow \infty$ 
coming from this $f(x)$ dominates over the correction coming from $h(x)$.

For the power law deviation, $1-n_s=2(p+1)/[p(p+2)N_e]$ is not ruled out only for $p=1$, for which we obtain $n_s=1-4/(3N_e)$, giving 
$0.9733$ for $N_e=50$, still within $2\sigma$ of the central value for Planck+WMAP. Already for $p=2$, the $N_e=50$ value is $n_s=0.9850$, more
than $3\sigma$ away from the central value.

We have given examples of just exponential and power law deviations from the asymptotic constant at $\varphi\rightarrow\infty$, but it is not hard to 
see that through appropriate choosing of the functions $f(x)$ and $g(x)$ we can get to any inflationary potential of the "new inflation" type (deviation from an 
asymptotic constant value for the potential), though generically (i.e., for functions with Taylor expansion at $x=1$), we get the Starobinsky model.

But in order to judge the naturalness of the model, from the point of view of the embedding into supergravity, the issue is how natural (i.e., simple
and likely to be obtained as an effective theory) is the superpotential? We will just give examples of the asymptotic behaviours. For the generic 
Starobinsky asymptotic behaviour in (\ref{genericstarob}), the asymptotic value of $W(\Phi)$ is 
\be
W(\Phi)=3\frac{\sqrt{\lambda}}{2}(n-2)M^3_{\rm Pl}\left[\frac{\Phi}{M_{\rm Pl}}-2n\sqrt{\frac{3}{2}}e^{-\sqrt{\frac{2}{3}}\frac{\Phi}{M_{\rm Pl}}}\right]\;,
\ee
and of $W(\Phi,S)$ is
\be
W(\Phi,S)=3(n-2)\sqrt{\lambda\a}M_{\rm Pl}^2S\Phi^{\frac{\a-1}{2}}\left[1-n \Phi^{-\sqrt{\frac{\a}{3}}}\right]\;,
\ee
and a similar one for the generalized Starobinsky model. For the asymptotic behaviour in (\ref{v0power}), the asymptotic value of $W(\Phi)$ is 
\be
W(\Phi)=M_{\rm Pl}\frac{\sqrt{V_0}}{2}\left[\frac{\Phi}{M_{\rm Pl}}-\frac{K}{2(p+1)}\left(\frac{M_{\rm Pl}}{\Phi}\right)^{p+1}\right]\;,
\ee
and of $W(\Phi,S)$ is 
\be
W(\Phi,S)=\sqrt{\a V_0}S\Phi^{\frac{\a-1}{2}}\left[1-K\left(\sqrt{\frac{\a}{2}}\ln \frac{\Phi}{M_{\rm Pl}}\right)^{-p}\right].
\ee
These behaviours are natural enough, though one would need to construct models that can interpolate between these behaviour and the 
Higgs behaviours at small field.

\section{Conclusions}

In this paper we have embedded Jordan frame Einstein+scalar models in minimal supergravity in two ways, using the constructions of 
\cite{Ketov:2014hya} and \cite{Roest:2013aoa}. We have seen that a generic $f(R)$ model can be written as a Einstein + dynamical scalar 
model, and reversely any inflationary potential can be written as $f(R)$, and embedded these models in ${\cal N}=1$ supergravity, and considered
the inflationary constraints on them. We have also seen that conformal inflation from the Higgs models can also be embedded in ${\cal N}=1$ 
supergravity, and are compatible with any kind of inflationary potential of "new inflation" type, and considered inflationary constraints on them.

We see that the simplicity and/or naturalness of inflationary models depend on the point of view, and how one intends to derive them from a more 
fundamental theory. An $f(R)$ model might look simple, yet the Einstein frame + scalar picture might be complicated, or reversely a simple 
Einstein frame picture can correspond to a complicated $f(R)$. A conformal inflation model might look simple, but the Einstein frame + scalar 
picture can be complicated. In the end, if these models come from minimal supergravity, we should ask how likely it is to have such a supergravity 
model as an effective action, perhaps coming from a more fundamental (e.g. string) theory? We can only answer this question in specific UV complete models,
but the goal of this paper was to set up the problem in a way that can be addressed, by relating various Jordan frame Einstein + scalar models to 
the Einstein frame and embedding them in minimal supergravity plus matter.

\section*{Acknowledgements}

I would like to thank Rogerio Rosenfeld for discussions and a critical reading of the manuscript.
My work is supported in part by CNPq grant 301709/2013-0 and FAPESP grant 2013/14152-7.

\bibliography{sugrainflation}
\bibliographystyle{utphys}

\end{document}